\documentclass[12pt]{article}
\usepackage{amssymb,amsmath,epsfig,cite}
 \allowdisplaybreaks

\begin{document}
\title{\bf Stability of Anisotropic Stellar Filaments}

\author{M. Zaeem-ul-Haq Bhatti \thanks{mzaeem.math@pu.edu.pk} and Z. Yousaf
\thanks{zeeshan.math@pu.edu.pk}\\
Department of Mathematics, University of the Punjab,\\
Quaid-i-Azam Campus, Lahore-54590, Pakistan.}

\date{}

\maketitle
\begin{abstract}
The study of perturbation of self-gravitating celestial cylindrical
object have been carried out in this paper. We have designed a
framework to construct the collapse equation by formulating the
modified field equations with the background of $f(R,T)$ theory as
well as dynamical equations from the contracted form of Bianchi
identities with anisotropic matter configuration. We have
encapsulated the radial perturbations on metric and material
variables of the geometry with some known static profile at
Newtonian and post-Newtonian regimes. We examined a strong
dependence of unstable regions on stiffness parameter which measures
the rigidity of the fluid. Also, the static profile and matter
variables with $f(R,T)$ dark source terms control the instability of
compact cylindrical system.
\end{abstract}
{\bf Keywords:} Relativistic system; Instability; Cylindrical system.\\
{\bf PACS:} 04.40.Cv; 04.40.Dg; 04.50.-h.

\section{Introduction}

One of the most important and remarkable research outcomes, over the
past few decades, is that our cosmos is expanding with an
accelerating rate. The roots of this discovery laid from the
observations of high red shift supernova Ia \cite{zf1}, which was
then reinforced by the cross comparison with the large scale
structure \cite{zf2} and cosmic microwave background radiations
\cite{zf3}. To explore such cosmic puzzle, one requires to introduce
a constituent in the cosmic matter distribution equipped with a huge
negative pressure gradient in most conventional gravitational
theory, i.e., general relativity (GR). Scientists dubbed this form
as dark energy (DE). Further, cosmological indications point flat
configurations of our universe with its approximate ratio of
constituents as $\frac{2}{3}$ dark energy and $\frac{1}{3}$ dark
matter. The need of the hour is to study the unknown and ambiguous
nature of DE as well as dark matter (DM). Due to this, many radical
various cosmic models have been suggested, like, a tiny no negative
and non zero cosmological constant, phantoms, quintessence,
Chaplygin gas, brane worlds DE and many more [See the review
articles \cite{zf4,zf5,zf6,zf7} and references therein].

Besides this, there is another challenging issue in relativistic
astrophysics, i.e., deep analysis of structure and formation of
compact stellar objects, like black holes, white dwarfs and neutron
stars. It is well-known that, on scales much smaller than the
horizon size, the DE fluctuations are insignificant \cite{zf10},
their influences on the relativistic fluid over densities evolution
are remarkable \cite{zf11}. This made the researchers to investigate
how DE alters the alluring phenomenon of stellar gravitational
collapse. Generally, it is believed that DE produces non-attractive
force on its environment, and thus it is expected that DE may
produce hindrances in the star collapse. Further, the burning
question is how DE mollify the dynamics of already established black
holes. It has been analyzed, recently, that black hole mass
decreases in the presence phantom energy accretion and approaches to
zero on reaching the Big Rip phase \cite{zf12}. Qadir et al. \cite{qadir1}
discussed various aspects of modified relativistic dynamics and
proposed that GR may need to be modified to resolve various
cosmological issues, like quantum gravity and the dark matter problem.

The effects of some well-consistent $f(R)$ corrections on the
formation of celestial structures lead to various phenomenological
consequences. In particular, gravity induced by extreme curvature
terms could yield significant deviations from that of GR. The first consistent accelerating universe model from $f(R)$
gravity was suggested in \cite{no1}. In this
scenario, number of results have been found (see Ref.\cite{zf14} for
a review), including restricted axially symmetric astronomical
solutions \cite{zf15}, cylindrical voids relevant models
\cite{zf16}, Raychaudhuri \cite{zf17} as well modified-Ellis
\cite{zf18} equations, existence of charged stellar bodies
\cite{zf19}, post-inflationary reheating phenomena \cite{zf20} or
the current accelerated cosmic expansion \cite{zf21}. The $f(R,T)$
gravity theory \cite{zf22} is one of the extensions of Einstein's
theory of gravity (GR) based on the coupling of relativistic
geometrical configurations with its matter content. In this theory,
the Lagrangian for EH action includes the extra degrees of freedom
along with trace of stress energy tensor.

It is worthy to stress that there exists wide range of papers that
considered the problem of Jeans analysis and referred subjects like
collapsing stellar systems and black holes in modified gravities.
For instance, the Jeans analysis for some celestial bodies has been
performed in $f(R)$ gravity by different astrophysicist \cite{zf23}.
Clifton \emph{et al.} \cite{zf24} discussed gravitational collapse
of relativistic spherical distributions and analyzed some of their
dynamical properties in the context of $f(R)$ gravity. However,
Sanders \cite{zf25} and Malekjani \emph{et al.} \cite{zf26}
addressed the issue of stellar collapse in modified Newtonian
dynamics. Martino \emph{et al.} \cite{zf27} studied the evolution of
spherical relativistic model model and found that modified gravity
corrections substantially affect evolutionary phases of huge cosmic
relativistic structures. Recently, Moraes \emph{et al.} \cite{n1}
presented some theoretical predictions of $f(R,T)$ gravity by
analyzing the existence and evolution static wormholes models.

Santos \cite{zf28} investigated the existence of spherical compact
stellar bodies in extended version of $f(R)$ gravity and found
widely different structure of relativistic neutron stars from that
found in GR. Ghosh and Maharaj \cite{zf29} examined collapse of
relativistic non-interacting particles in $f(R)$ gravity and
determined relatively less unstable matter distribution in the
mysterious dark cosmos. Cembranos \emph{et al.} \cite{zf30} also
performed such analysis and established that gravity predicted
$f(R)$ gravity supports more compact stellar system configurations.
Alavirad and Weller \cite{zf31} studied the effects of finite
logarithmic $f(R)$ terms in the structure formation of stellar
interior and observed some attractive observational outcomes
differing from the dynamics induced by GR. Sebastiani \emph{et al.}
\cite{zf32} found both stable as well as unstable black holes
configurations in the presence of gravity induced by some $f(R)$
models. various dynamical features of wormholes \cite{ol6,il1}, black
holes \cite{ol7, ol8} and curvature singularity
\cite{Arbuzova:2010iu, 6l} in the relativistic celestial bodies have
been explored in $f(R)$ gravity.

Sharif with his research fellows \cite{zf33} examined various
stability islands for some relativistic compact matter
distributions. Further, they also found role of some matter
variables in this context \cite{zf34}. Farinelli \emph{et al.}
\cite{zf35} numerically solved the extended versions of Lane-Emden
equation for relativistic systems and determined that $f(R)$ gravity
is probably to host massive stable compact objects. Bhatti and
Yousaf \cite{zf36} checked the existence of various interesting
configurations of stellar objects and encountered wide variety of
huge celestial homogeneous objects in $f(R)$ gravity. Recently,
Yousaf \emph{et al.} \cite{zf37} explored that how modified gravity
affects the initial regular cosmic environment for the collapse of
compact celestial systems.

The analysis of stability issue has been motivated from some
gravitational tests obtained from pulsar-timing experiments. The
pioneer work in the development of dynamical instability was done by
Chandrasekhar \cite{zf38} by giving a relation about the stability
of spherical star with the help of adiabatic index, $\Gamma_1$. The
interesting feature of $\Gamma_1$ is that it provides enough
information about the stiffness of the relativistic collapsing
interiors. The non-adiabatic phenomenon of spherical gravitational
collapse has been explored by \cite{zf39} and confronted that
dissipation caused by heat radiations boost up the unstable phases
of the evolving stars at both Newtonian (N) and post-Newtonian (pN)
epochs. Chan \emph{et al. }\cite{zf40} investigated that tiny value
of anisotropy in the pressure configurations could dramatically
disturb the stable phases of the self-gravitating spherical
interior. Herrera and Santos \cite{zf41} presented a detailed review
on the importance of anisotropic pressure in the modeling of
collapsing stellar interiors. Sharif and his collaborators
\cite{zf42} investigated the relationship between fluid thickness
and matter parameters for various cosmological compact models.
Yousaf and Bhatti \cite{zf43} identified instability constraints for
the stability of expansion-free self-gravitating compact objects in
the weak-field approximations of $f(R,T)$ gravity.

Here, we study the role of stiffness parameter and $f(R)$ dark
source terms to understand the unstable epochs of anisotropic
relativistic interiors in $f(R,T)$ gravity. The paper is organized
as follows. In section one, we provide $f(R,T)$ field equations,
junction conditions as well as dynamical equations. Section two
provides well-known radial configuration of perturbation. We then
employ this over various fundamental equations of our dynamical
system. In section three, we use Harrison-Wheeler state equation to
relate perturbed pressure gradient with the systems energy density,
after that we find $f(R,T)$ collapse equation. Section four explores
the instability regions of cylindrical anisotropic system with N pN
limits. We conclude our results in the last section.

\section{$f(R,T)$ Theory and Stellar Filament}

Though considered in many other astrophysical frameworks, the
concept of $f(R)$ gravity received considerable attention mainly due
to the the reason that it could presents plausible description to
the alluring phenomenon of accelerating universe expansion
\cite{zf44}. The main idea in $f(R,T)$ gravity is to replace the
tiny value of cosmological constant in usual EH with an algebraic
generic expression of $R$ and $T$, where $R$ is the Ricci scalar,
while $T$ is the trace of energy momentum tensor. It is given as
\cite{zf22}
\begin{equation}\label{1}
S_{f(R,T)}=\int d^4x\sqrt{-g}[f(R,T)+L_M],
\end{equation}
where $g$ is the metric trace, while $L_M$ is indicating the
presence of matter Lagrangian. It worthy to mention that the
dynamical equations of motions in $f(R,T)$ gravity theory is
directly related the matter contents contribution, therefore,
relativistic astrophysicists can get any particular configurations
of equations by choosing any form $L_M$. Here, we consider $L_M=\mu$
(where $\mu$ stands for system energy density). Now, we vary the
above action with $g_{\mu\beta}$, after some manipulation, we obtain
the following distributions of $f(R,T)$ field equations (for details
\cite{zf27})
\begin{equation}\label{2}
{G}_{\mu\beta}={{T}_{\mu\beta}}^{\textrm{eff}},
\end{equation}
where
\begin{align*}\nonumber
{{T}_{\mu\beta}}^{\textrm{eff}}&=\left[(1+f_T(R,T))T^{(m)}_{\mu\beta}-\mu
g_{\mu\beta}f_T(R,T)
-\left(\frac{f(R,T)}{R}-f_R(R,T)\right)\frac{R}{2}\right.+\\\nonumber
&\left.+\left({\nabla}_\mu{\nabla}_
\beta+g_{\mu\beta}{\Box}\right)f_R(R,T)\right]\frac{1}{f_R(R,T)}.
\end{align*}
Here, ${G}_{\mu\beta}$ and ${{T}_{\mu\beta}}^{\textrm{eff}}$ are
Einstein and effective $f(R,T)$ energy-momentum tensors,
respectively. Further, $\Box$ is the de-Alembert's operator
expressed by means of covariant derivation operator as
$\nabla_\mu\nabla^\mu$ while subscripts $T$ and $R$ indicate
derivations with respect to $T$ and $R$ of the corresponding
quantities.

We take a cylindrical self-gravitating system filled with locally
anisotropic fluid configurations. We assume that the evolution
phases of this system is separated by a $3D$ timelike surface. This
surface is symbolized with $\Sigma$. This boundary differentiates
our cylindrical manifold, $\mathcal{V}$, into couple of portions, in
which the exterior one is denoted with the help of $+$ sign, while
$-$ sign indicates configuration of interior manifold. The
$\mathcal{V}^-$ manifold can be mentioned through the following
diagonal and non-rotating metric
\begin{equation}\label{3}
ds^2_-=A^2dt^{2}-B^2dr^{2}-C^2d\phi^{2}-D^2dz^2.
\end{equation}
where metric coefficients are the functions of $t$ and $r$. For the
sake of simplicity, we are considering $D$ to be unity. Such type of
assumption has been taken by many researchers \cite{1r, v33a, v33b}.
The spacetime for $\mathcal{V}^+$ is \cite{v34}
\begin{equation}\label{4}
ds^2_+=-e^{2(\gamma-\upsilon)}(d\nu^2-d\rho^2)+e^{-2\upsilon}\rho^2d\phi^2+e^{2\upsilon}d{z}^2,
\end{equation}
where $\gamma$ and $\upsilon$ are the functions of $\nu$ and $\rho$,
while the coordinates are numbered as
$x^{\beta}=(\nu,~\rho,~\phi,~z)$. The corresponding vacuum field
equations provide
\begin{align}\label{k1}
&\rho(\upsilon_\nu^2+\upsilon_\rho^2)=\frac{f-Rf_R}{2f_R}e^{2(\gamma-\upsilon)},\\\label{k2}
&2\upsilon_\nu\upsilon_\rho \rho=\gamma_\nu,\\\label{k3}
&\upsilon_{\nu\nu}-\frac{\upsilon_{\rho}}{\rho}-\upsilon_{\rho\rho}=\frac{e^{2(\gamma-\upsilon)}}{4\rho}
\left(\frac{f-Rf_R}{f_R}\right)\left\{\rho
e^{-4\upsilon}+\frac{e^{2\gamma}}{\rho}\right\},
\end{align}
where subscripts $\rho$ and $\nu$ show partial differentiations with
respect to $\rho$ and $\nu$, respectively. To calculate the $f(R,T)$
field equations, we choose  the following form of the usual
energy-momentum tensor \cite{v35}
\begin{equation}\label{5}
T^-_{\mu\beta}=(\mu+P_{r})V_{\mu}V_{\beta}-P_r
g_{\mu\beta}+(P_{z}-P_{r})S_{\mu}S_{\beta}+ (P_\phi-P_{r})K_{\mu}
K_{\beta},
\end{equation}
where $P_{z},~P_{\phi}$ and $P_{r}$ are pressure gradients along
$z,~\phi$ and $r$ directions, respectively. Moreover,
$V_{\beta}=A\delta^{0}_{\beta}$ is the four-velocity, while
$S_{\beta}={\delta}^{3}_{\beta},~ K_{\beta}=C{\delta}^{2}_{\beta}$
are four-vectors. We now consider that our system is configuring in
the comoving coordinate system. In this system, above four vectors
obey
\begin{equation*}
V^{\beta}V_{\beta}=-1,\quad
K^{\beta}K_{\beta}=S^{\beta}S_{\beta}=1,\quad
V^{\beta}K_{\beta}=S^{\beta}K_{\beta}=V^{\beta}S_{\beta}=0.
\end{equation*}
The expansion scalar $\Theta=V^{\mu}_{~;\mu}$ for our cylindrically
interior turns out to be
\begin{equation}\label{6}
\Theta=\frac{1}{A}\left(\frac{\dot{B}}{B} +\frac{\dot{C}}{C}\right),
\end{equation}
where over dot means $\frac{\partial}{\partial t}$.

The $f(R,T)$ field equations (\ref{2}) for the cylindrically
symmetric spacetime (\ref{3}) yield
\begin{align}\label{7}
&G_{00}=\frac{A^2}{f_R}\left[\mu-\left(f_R-\frac{f}{R}\right)\frac{R}{2}+\psi_{tt}\right],\quad
G_{01}=\psi_{tr},\\\label{8}
&G_{11}=\frac{B^2}{f_R}\left[P_r+(P_r+\mu){f_T}
+\left(f_R-\frac{f}{R}\right)\frac{R}{2}+\psi_{rr}\right],\\\label{9}
&G_{22}=\frac{C^2}{f_R}\left[P_\phi+(P_\phi+\mu){f_T}+\left(f_R-\frac{f}{R}
\right)\frac{R}{2}+\psi_{\phi\phi}\right],\\\label{10}
&G_{33}=\frac{1}{f_R}\left[P_z+(P_z-\mu){f_T}+\left(f_R-\frac{f}{R}\right)\frac{R}{2}+\psi_{zz}\right],
\end{align}
where
\begin{align}\label{11}
&\psi_{tt}=\frac{\partial_{rr}f_{R}}{B^2}-
\left(\frac{\dot{B}}{B}+\frac{\dot{C}}{C}\right)\frac{\partial_tf_R}{A^2}
+\left(\frac{C'}{C}-\frac{B'}{B}\right)\frac{\partial_rf_{R}}{B^2},\\\label{12}
&\psi_{tr}=\frac{1}{f_R}\left(\partial_r\partial_tf_{R}
-\frac{\dot{B}}{B}\partial_rf_{R}-\frac{A'}{A}\partial_tf_{R}\right),\\\label{13}
&\psi_{rr}=\frac{\partial_t\partial_tf_R}{A^2}-\left(
\frac{\dot{A}}{A}-\frac{\dot{C}}{C}\right)\frac{\partial_tf_{R}}{A^2}
-\left(\frac{A'}{A}+\frac{C'}{C}\right)\frac{\partial_rf_{R}}{B^2},\\\label{14}
&\psi_{\phi\phi}=\frac{\partial_{rr}f_{R}}{B^2}+\frac{\partial_{tt}f_{R}}{A^2}
+\left(\frac{\dot{B}}{B}-\frac{\dot{A}}{A}\right)
\frac{\partial_tf_{R}}{A^2}
+\left(\frac{B'}{B}-\frac{A'}{A}\right)\frac{\partial_rf_{R}}{B^2},\\\label{15}
&\psi_{zz}=\frac{\partial_{tt}f_{R}}{A^2}-
\left(\frac{\dot{A}}{A}-\frac{\dot{B}}{B}+\frac{\dot{C}}{C}\right)\frac{\partial_tf_{R}}{A^2}
-\frac{\partial_{rr}f_{R}}{B^2}+\left(\frac{B'}{B}-\frac{A'}{A}-\frac{C'}{C}\right)\frac{\partial_rf_{R}}{B^2}.
\end{align}
where prime indicates $\frac{\partial}{\partial r}$ operator. The
Ricci invariant for the non-ideal cylindrical metric with $f(R,T)$
gravity are found as
\begin{eqnarray}\nonumber
R(t,r)&=&\frac{2}{B^2}\left[\frac{A''}{A}+\frac{C''}{C}+\frac{A'}{A}
\left(\frac{C'}{C}-\frac{B'}{B}\right)-\frac{B'C'}{BC}\right]\\\label{19}
&&-\frac{2}{A^2}\left[\frac{\ddot{B}}{B}+\frac{\ddot{C}}{C}-\frac{\dot{A}}{A}
\left(\frac{\dot{B}}{B}+\frac{\dot{C}}{C}\right)+\frac{\dot{B}\dot{C}}{BC}\right].
\end{eqnarray}

\section{Dynamics}

In this section, we shall perform dynamical investigations of a
stellar filaments coupled with anisotropic matter content in
$f(R,T)$ gravity. In this setting, we shall compute the
hydrodynamical equation for the system which is undergoing
oscillations. We shall also find instability epochs of the perturbed
cylindrical body.

\subsection{Hydrodynamics}

Now, we shall formulate dynamical equations with the help of
energy-momentum tensor divergence. This will assists us to analyze
dynamical phases of cylindrical relativistic stellar bodies with
$f(R,T)$ background. It is worthy to note that in this modified
gravity theory, the divergence of energy momentum tensor is
non-zero. In this case, we have the following configurations of the
divergence \cite{n1}
\begin{align}\label{16}
\nabla^{\mu}T_{\mu\beta}=\left[
(\Theta_{\mu\beta}+T_{\mu\beta})\nabla^\mu{\ln}f_T
+\nabla^\mu\Theta_{\mu\beta}-\frac{1}{2}g_{\mu\beta}\nabla^\mu{T}\right]\frac{f_T}{1-f_T}.
\end{align}
For our locally anisotropic cylindrical system, the above equation
gives
\begin{align}\nonumber
&\dot{\mu}\left(\frac{1+2f_T}{f_R(1+f_T)}\right)
-\frac{B\dot{B}}{A^2f_R}(1+f_T)(\mu+P_r)-\frac{\mu}{f_R}\partial_tf_R
-\frac{C\dot{C}}{A^2f_R}(1+f_T)(\mu+P_\phi)\\\label{17}
&+\frac{\partial_tT}{2(1+f_T)}+\frac{2\mu}{1+f_T}\partial_tf_T+D_0=0,
\\\nonumber
&\frac{P'_{r}}{f_R}
-\frac{AA'}{A^2f_R}(1+f_T)(\mu+P_r)+\frac{P_{r}}{f_R}\left\{\partial_rf_T
-\frac{(1+f_T)\partial_rf_R}{f_R}\right\}
+\frac{\mu'}{f_R}+\frac{\mu}{f_R}\\\nonumber
&\times\left\{\partial_rf_T-\frac{f_T\partial_rf_R}{f_R}\right\}
+\frac{1}{(1+f_T)}\left\{\left(\frac{T'}{2}+\mu'\right)f_T-(\mu-P_r)
\partial_rf_T\right\}+\frac{C'}{CB^2f_R}\\\label{18}
&\times(P_r-P_\phi)(1+f_T)+D_1=0,
\end{align}
where $D_0$ and $D_1$ are higher degrees of freedom offered by
$f(R,T)$ theory.These are the functions of both temporal and radial
coordinates. The dark source terms are energy variations corrections
along with the time and adjacent surfaces in the cylindrical
self-gravitating celestial objects, respectively. These are
calculated and written in Appendix \textbf{A}.

Thorne \cite{v35a} defined the C-energy for cylindrically symmetric
spacetime as follows
\begin{equation*}
\tilde{E}(t,r)={m}(t,r)=\frac{1}{8}(1-l^{-2}
{\nabla}^{\alpha}{\tilde{r}}{\nabla}_{\alpha}{\tilde{r}}).
\end{equation*}
The circumference radius, $\rho$, specific length, $l$, and areal
radius $\tilde{r}$ of the cylindrical geometry obey following
relations
\begin{equation*}
\rho^2={\xi_{(1)a}}{\xi^a_{(1)}},\quad
l^2={\xi_{(2)a}}{\xi^a_{(2)}},\quad {\tilde{r}}=\rho{l}.
\end{equation*}
where $\xi_{(1)}=\frac{\partial}{\partial{\phi}}$ and
$\xi_{(2)}=\frac{\partial}{\partial{z}}$ are the Killing vectors in
cylindrical system. The specific energy in our cylindrical geometry
is found as
\begin{equation}\label{20}
m(t,r)=\frac{l}{8}\left(1-\frac{C'^2}{B^2}+\frac{\dot{C}^2}{A^2}\right).
\end{equation}

Now, we define couple of well-known operators, i.e., proper and
radial derivative operators. This will helps us to find the
cylindrical mass function variations among its adjacent surfaces.
These operators are
\begin{eqnarray}\label{21}
D_{J}\equiv\frac{1}{A} \frac{\partial}{\partial t},\quad
D_{C}\equiv\frac{1}{C'}\frac{\partial}{\partial r}.
\end{eqnarray}
Using Eqs.(\ref{20}) and (\ref{21}), we get
\begin{eqnarray}\label{22}
{\mathrm{E}}\equiv\frac{C'}{B}=\sqrt{1-\frac{8m}{l}+U^{2}}.
\end{eqnarray}
where $U$ is the relativistic system velocity and is given as
\begin{eqnarray}\label{23}
U=D_{J}C=\frac{\dot{C}}{A}.
\end{eqnarray}
Equations (\ref{7}), (\ref{8}) and (\ref{20})-(\ref{23}) provide
\begin{align}\label{24}
D_{C}m&=\frac{l}{4}\left[\mu-
\left(f_R-\frac{f}{R}\right)\frac{R}{2}+\psi_{tt}-
\frac{\psi_{tr}}{BA}\frac{U}{{\mathrm{E}}}\right]\frac{C}{f_{R}}.
\end{align}
This equation describes total energy variation among neighboring
anisotropic fluid surfaces. It is evident that the first three
mathematical terms on the right hand side in the above equation are
coming due to $f(R,T)$ system energy density corrections. The last
term $(\psi_{tr}/BA)(U/{\mathrm{E}})$  contributes to affect the
evolutionary mechanism of the cylindrical system due to its
non-attractive behavior. It is note worthy that $\psi_{tr}>0$ and
for collapsing models, $U$ must be negative, therefore
$(\psi_{tr}/BA)(U/{\mathrm{E}})<0$ in the above equation. This will
make $-(\psi_{tr}/BA)(U/{\mathrm{E}})$ to be greater than zero. The
solution of Eq.(\ref{24}) is
\begin{align}\label{25}
m=\frac{1}{4}\int^C_{0}\left[\left\{\mu-
\left(f_R-\frac{f}{R}\right)\frac{R}{2}+\psi_{tt}-
\frac{\psi_{tr}}{BA}\frac{U}{{\mathrm{E}}}\right\}\frac{l}{f_{R}}\right]dC.
\end{align}

For the smooth matching of the interior and exterior geometry over
the boundary surface $\Sigma$, we will use the Darmois \cite{v36} as
well as Senovilla \cite{sano1} junction conditions. We consider
timelike hypersurface for which, we impose $r=$constant in
Eq.(\ref{3}) and $\rho(\nu)$ in Eq.(\ref{4}). In this context, the
first fundamental form on $\Sigma$, provide
\begin{align}\label{26a}
&d\tau\overset{\Sigma}=e^{2\gamma-2\upsilon}\sqrt{1-\left(\frac{d\rho}{d\nu}\right)^2}d\nu=Adt,\\\label{26b}
&C\overset{\Sigma}=e^{\upsilon},\quad e^{\upsilon}=\frac{1}{\rho},
\end{align}
with $1-\left(\frac{d\rho}{d\nu}\right)^2>0$. The second fundamental
form gives us the following set of equations
\begin{align}\label{26c}
&e^{2\gamma-2\upsilon}[\nu_{\tau\tau}\rho_{\tau}-\rho_{\tau\tau}\nu_{\tau}
-\{\rho_\tau(\gamma_\nu-\upsilon_\nu)+\nu_\tau(\gamma_\rho-\upsilon_\rho)\}(\nu_\tau^2-\rho_\tau^2)]
\overset{\Sigma}=\frac{-A'}{AB},\\\label{26d}
&e^{2\upsilon}(\rho_\tau\upsilon_{\nu}+\nu_\tau\upsilon_\rho)\overset{\Sigma}=\frac{CC'}{B},\quad
\rho_\tau\upsilon_\nu+\nu_\tau\upsilon_\rho\overset{\Sigma}=\frac{\nu_\tau}{\rho}.
\end{align}
Using Eqs.(\ref{26a})-(\ref{26d}) and field equations, we get
\begin{align}\label{27}
& P_r^{\textrm{eff}}\overset{\Sigma}=0.
\end{align}
Further, we have
\begin{align}\label{26}
R|_-^+&=0,\quad f_{,RR}[\partial_\nu R|_-^+=0,\quad f_{,RR}\neq0.
\end{align}
The constraint mentioned in Eq.(\ref{27}) is due to Darmois junction
conditions. Equation (\ref{26}) is required to be obeyed over the
hypersurface due to modified gravity which affirms the continuity of
Ricci curvature invariant over $\Sigma$ even for matter thin shells.
By making use of Eqs.(\ref{27}) and (\ref{26}), we have
\begin{align}\label{26}
P_r\overset{\Sigma}=\frac{1}{1+f_T}\left(\frac{f-Rf_R}{2}\right).
\end{align}
This equations shows that radial pressure on the boundary surface is
non-zero and depends on the dark source terms coming from $f(R,T)$
gravity with constant $R$ and $T$ background. Thus, modified gravity
corrections are directly linked with the flux of momentum of the
gravitational wave emerging from the relativistic cylindrical
system.

For the presentation of cosmological and theoretical well-consistent
$f(R,T)$ theory, the selection of its generic function holds
fundamental importance. In this study, we take specific class of
$f(R,T)$ models given as follows
\begin{equation}\label{29}
f(R,T)=f_1(R)+f_2(R)f_3(T).
\end{equation}
The model of this type of configurations are originated from the
explicit non-minimal coupling of curvature and matter. We now study
$f(R,T)$ power law type model  mentioned as follows
\begin{equation}\label{30}
f(R)=R+\lambda R^2T^2.
\end{equation}
Such functional form meet with the Lagrangian form of $f(R,T)$
gravity observed in Eq.(\ref{29}). In the above equation, we have
taken $f_1(R)=R,~f_2(R)=R^2$ along with $f_3(T)=T^2$. By considering
$\lambda\rightarrow0$, all GR solutions can be maintained.

A detailed review on the viability of $f(R,T)$ models has been presented in \cite{1c2}.
We have assumed $f(R,T)$ model mentioned in Eq.(\ref{30}),
in which the value of $\lambda$ should indeed be small in order to
satisfy the Solar system tests. If one takes $f(R,T)=R+\lambda T$,
then one can obtain dynamics identical with that of GR. Thus, the
simple case $f(R,T)=R+\lambda T$ is fully equivalent with standard
GR, after rescaling $\lambda$. The value of $\lambda$ should be
small enough, i.e. $\lambda < < 1$. The $f(R,T)$ theory with this
model is just general relativity with strange dynamical cosmological
constant $f(T)$ (covariant!) or if one wish with strange matter
coupling. Further, the gravitational dynamics induced by $R+2\lambda T$
gravity model give results comparable with that gravitational model
having an effective cosmological constant. i.e.,
$\Lambda_{eff}\propto H^2$ (where $H$ is a Hubble parameter). In
this standpoint, the gravitational coupling would behave as an
effective time dependent coupling, thereby providing
$G_{eff}=G\pm\lambda$. The study of the cosmological perturbations
of $f(R,T)=R+\lambda R^2T^2$ model has not been performed yet, and
this would give another set of constraints on $\lambda$.

\subsection{Oscillations}

In this section, we shall proceed our analysis by considering a
well-known radial perturbation method. We shall employ this
technique over $f(R,T)$ field and dynamical equations. In this
scheme, the quantity $\alpha$ is known as perturbation parameter
that would control the impact of fluctuations within the cylindrical
anisotropic system. We would consider effects upto $O(\alpha)$. It
is worthy to stress that $\alpha\in(0,1)$ in our examination. This
perturbation scheme was brought in by Herrera \textit{et al.}
\cite{her1a}. The ``static" quantities of the corresponding
expressions are represented with zero subscript. Here, we take
perturbation with the assumption that all structural variables of
the systems in $f(R,T)$ gravity are dependent upon the same time
dependent function, i.e., $\eta(t)$. The perturbation technique is
\begin{eqnarray}\label{31}
X(t,r)&=&X_0(r)+{\alpha}\eta(t)x(r),\\\label{35}
Y(t,r)&=&Y_{0}(r)+{\alpha}{\bar{Y}}(t,r),\\\label{40}
f(t,r)&=&R_0(1+T_0^2{\lambda}R_0)+\alpha
\eta(t)e(r)(1+2R_0{\lambda}T_0^2),\\\label{41}
f_R(t,r)&=&(1+2R_0{\lambda}T_0^2)+2{\alpha}\eta(t)
e(r){\lambda}T_0^2,\\\label{42}
\Theta(t,r)&=&0+\alpha{\bar{\Theta}}(t,r),
\end{eqnarray}
where Eq.(\ref{31}) shows oscillations of metric variables while
oscillations of matter variables are represented in Eq.(\ref{35}).
The non-static perturbed configurations of Eq.(\ref{19}) is
\begin{align*}
&-e\eta=\left(\frac{b}{B_0}+\frac{c}{r}\right)\frac{2\ddot{\eta}}{A_0^2}
+\eta\left[\frac{4R_0b}{B_0^3}
-\frac{2}{B_0^2}\left\{\frac{c''}{r}-\frac{aA''_0}{A_0^2}
+\frac{a''}{A_0}-\frac{A'_0}{A_0}\left\{\left(\frac{b}{B_0}\right)'
\right.\right.\right.\\\nonumber
&-\left.\left.\left.\left(\frac{c}{r}\right)'\right\}+
\left(\frac{1}{r}-\frac{B'_0}{B_0}\right)\left(\frac{a}{A_0}\right)'
-\left(\frac{c}{r}\right)'\frac{B'_0}{B_0}
-r^{-1}\left(\frac{b}{B_0}\right)' \right\}\right],
\end{align*}
while its static form is
\begin{equation*}
R_0(r)=2\left[\frac{A'_0}{A_0}
\left(\frac{1}{r}-\frac{B'_0}{B_0}\right)+\frac{A''_0}{A_0}-\frac{B'_0}{B_0r}\right]\frac{1}{B_0^2}.
\end{equation*}
The static form of the locally anisotropic cylindrical $f(R,T)$
field equations (\ref{7})-(\ref{10}), with $C_0=r$, are given as
follows
\begin{align}\label{43}
&(1+2{\lambda}R_0T_0^2)\left(\frac{B'_0}{rB_0}\right)=rB_0^2\left[\mu_0-\frac{\lambda}{2}
R_0^2T_0^2+\overset{(S)}{\psi_{tt}}\right],
\\\label{44}
&(1+2{\lambda}R_0T_0^2)\left(\frac{A'_0}{A_0}\right)=rB^2_0\left[(P_{r0}
+\mu_0)(1+2{\lambda}R_0^2T_0)+\frac{\lambda}{2}R_0^2T_0^2+\overset{(S)}{\psi_{rr}}
\right],\\\nonumber
&(1+2{\lambda}R_0T_0^2)\left(\frac{B'_0}{B^2_0}+\frac{A''_0}{B_0A'_0}\right){A'_0}
=B_0A_0\left[(P_{\phi0}+\mu_0)(1+2{\lambda}R_0^2T_0)\right.\\\label{45}
&
+\left.\frac{\lambda}{2}R_0^2T_0^2+\overset{(S)}{\psi_{\phi\phi}}\right],\\\nonumber
&(1+2{\lambda}
R_0T_0^2)\left\{\left(\frac{A'_0}{A_0r}+\frac{A''_0}{A_0}\right)-\frac{B'_0}{B_0}
\left(\frac{1}{r}+\frac{A'_0}{A_0}\right)\right\}
={B_0^2}\left[(P_{z0}-\mu_0)(1+2{\lambda}\right.\\\label{46}
&\left.\times R_0^2T_0)
+\frac{\lambda}{2}R_0^2T_0^2+\overset{(S)}{\psi_{zz}}\right],
\end{align}
where overset index $(S)$ points out static form of the respective
$f(R,T)$ dark component terms. These quantities are computed as
follows
\begin{align}\nonumber
\overset{(S)}{\psi_{tt}}&=\frac{2
T_0\lambda}{B_0^2}\left[2T'_0R'_0+R''_0T_0+2T''_0R_0-\frac{2T'^2_0}{T_0}R_0
+\left(\frac{B'_0}{B_0}-\frac{1}{r}\right)(R'_0T_0+2T_0'R_0)\right],\\\nonumber
\overset{(S)}{\psi_{rr}}&=-\frac{2{T_0}\lambda}{B_0^2}(R'_0T_0
+2T'_0R_0)\left(\frac{1}{r}+\frac{A'_0}{A_0}\right),\\\nonumber
\overset{(S)}{\psi_{\phi\phi}}&=\frac{2\lambda{T_0}}{B_0^2}\left[
2T'_0R'_0+R''_0T_0+2T''_0R_0+2\frac{T'^2_0}{T_0}R_0
-\left(\frac{A_0'}{A_0}-\frac{B'_0}{B_0}\right)(T_0R'_0\right.+\left.2T'_0R_0)\right],\\\nonumber
\overset{(S)}{\psi_{zz}}&=\frac{2\lambda
T_0}{B_0^2}\left[(R'_0T_0+2T'_0R_0)\left(\frac{B'_0}{B_0}-\frac{1}{r}-\frac{A'_0}{A_0}\right)-
2T'_0R'_0-R''_0T_0-2T''_0R_0\right.-\left.2\frac{T'^2_0}{T_0}R_0\right].
\end{align}
The non-static perturbed form of (00), (11) and $(22)$ components of
$f(R,T)$ field are
\begin{align}\label{47}
&\bar{\mu}=\chi_2\eta,\\\nonumber &\bar{P_r}+\bar{\mu}=\eta\chi_3
-\frac{1}{(1+2{\lambda}R_0^2T_0)}\left[{\overset{(P_1)}{\psi_{tt}}\dot{\eta}}
+2\ddot{\eta}\left\{\overset{(P_2)}{\psi_{tt}}
+\frac{c}{rA_0^2}(1+2{\lambda}R_0T_0^2)\right\}\right.\\\label{48}
&\times\left.\left(\frac{b}{B_0^2}+T_0^2e\lambda\right)\right],\\\label{49}
&\bar{P_\phi}+\bar{\mu}=\eta\chi_5-\ddot{\eta}
\left[\frac{b}{B_0A_0^2}(1+2R_0{\lambda}T_0^2)+\overset{(P_2)}{\psi_{\phi\phi}}
\right]\frac{1}{(1+2{\lambda}R_0^2T_0)},
\end{align}
where $\chi_i's$ are static $f(R,T)$ model dependent dark origin
terms. All the impact of relativistic time variable in the dynamical
evolution of cylindrically symmetric field equations have been
successfully endowed in a single variable $\eta(t)$. The terms
$\chi_2,\chi_3$ and $\chi_5$ contain combinations of static metric
parameters corresponding to (\ref{7})-(\ref{9}) $f(R,T)$ field
equations, respectively. These quantities are described in Appendix
\textbf{A}. We see that Eq.(\ref{17}) is trivially obeyed under
perturbation, however, Eq.(\ref{18}) provides the following
dynamical equation
\begin{align}\nonumber
&\frac{P'_{r0}}{(1+2R_0\lambda T_0^2)}+\frac{2P_{r0}\lambda}{(1+2
R_0\lambda
T_0^2)}\left[R_0(R_0T'_0+2T_0R'_0)-\frac{(2R_0T'_0+T_0R'_0)T_0}{(1+2
R_0\lambda T_0^2)}\right.\\\nonumber &\left.\times(1+2R_0^2\lambda
T_0)\right]\left[\frac{a}{A_0}+\frac{b}{B_0}+\frac{2\lambda e
T_0^2}{(1+2R_0{\lambda}T_0^2)}\right]+\frac{\mu'_0}{(1+2
R_0{\lambda}T_0^2)}+\frac{2\mu_0{\lambda}}{(1+2
R_0{\lambda}T_0^2)}\\\nonumber
&\times\left[(R_0T'_0+2T_0R'_0)R_0-\frac{2
T_0^2{\lambda}R_0^2(2R_0T'_0+T_0R'_0)}{(1+2
R_0{\lambda}T_0^2)}\right]-2
R_0{\lambda}\frac{(2R_0T'_0+2T_0R'_0)}{(1+2R_0^2{\lambda}T_0)}\\\nonumber
&\times(\mu_0-P_{r0})+\frac{2R_0^2{\lambda}T_0}{(1+2
R_0{\lambda}T_0^2)}\left(\frac{T'_0}{2}+\mu_0'\right)+\frac{r(1+2
R_0^2{\lambda}T_0)}{B_0^2(1+2
R_0{\lambda}T_0^2)}(P_{r0}-P_{\phi0})\\\label{50}
&-\frac{A_0A'_0}{(1+2R_0{\lambda}T_0^2)B_0^2}(P_{r0}+\mu_0)(1+2
R_0^2{\lambda}T_0)+D_{1S}=0,
\end{align}
where $D_{1S}$ indicates that $D_1$ is evaluated under static form
of perturbation method. Its value is
\begin{align}\nonumber
D_{1S}&=\overset{(S)}{\psi_{rr,r}}-\frac{A_0A'_0}{(1+2 R_0\lambda
T_0^2)B_0^2}\left(\overset{(S)}{\psi_{tt}}+\overset{(S)}{\psi_{rr}}\right)+
(R_0^2T_0^2)'\frac{\lambda}{2}
+\frac{r\left(\overset{(S)}{\psi_{rr}}+\overset{(S)}{\psi_{\phi\phi}}\right)}{(1+2
R_0\lambda T_0^2)B_0^2}.
\end{align}
The hydrostatic as well as non-hydrostatic distributions of C-energy
function (\ref{20}) is
\begin{align}\label{51}
&m_0=\frac{l}{8}-\frac{l}{8B_0^2},\quad
\bar{m}=\frac{\eta}{B_0^2}\left(\frac{b}{B_0}-c'\right)\frac{l}{4},
\end{align}
The mathematical interaction between radial components $b$ and $B_0$
can be found from Eq.(\ref{8}) as
\begin{eqnarray}\nonumber
\frac{b}{B_0}=\left(\psi_{tr}+\frac{c'}{r}-\frac{cA_0'}{rA_0}\right)r.
\end{eqnarray}
Equations (\ref{17}) and (\ref{18}) after using
Eqs.(\ref{31})-(\ref{41}) give
\begin{align}\label{52}
&\bar{\dot{\mu}}+X_1(r)\dot{\eta}=0,
\\\nonumber
&\frac{\bar{P'_r}}{(1+2 R_0^2\lambda
T_0)}+\frac{2\lambda\bar{P_r}}{(1+2 R_0^2\lambda
T_0)}\left[(R_0T'_0+2T_0R'_0)R_0-(2R_0T'_0+T_0R'_0)T_0\right.\\\nonumber
&\times\left.\frac{(1+2R_0^2\lambda T_0)}{(1+2 R_0\lambda
T_0^2)}\right]+\frac{\bar{\mu'}}{(1+2R_0^2\lambda
T_0)}-\frac{A_0}{B^2_0}(\bar{\mu}+\bar{P_r}) A'_0
+\frac{2\bar{\mu}\lambda}{(1+2 R_0^2\lambda
T_0)}\left[(2T_0\right.\\\nonumber
&\times\left.R'_0+T'_0R_0)R_0-\frac{2R_0^2\lambda T_0^2}{(1+2
R_0\lambda T_0^2)}\right]+\frac{2 R_0^2\lambda
T_0\bar{\mu'}}{(1+2R_0^2\lambda T_0)}-2 R_0\lambda
\frac{(2T_0R'_0+R_0T'_0)}{(1+2R_0^2\lambda T_0)}\\\nonumber
&\times(\bar{\mu}-\bar{P_r})-\frac{A_0A'_0}{B_0^2(1+2R_0\lambda
T_0^2)}
\left(\dot{\eta}\overset{(P_1)}{\psi_{rr}}+\ddot{\eta}\overset{(P_2)}{\psi_{rr}}\right)
+\frac{r}{B_0^2}(\bar{P_r}-\bar{P_\phi})\frac{(1+2 R_0^2\lambda
T_0)}{(1+2R_0\lambda T_0^2)}\\\label{53} &+\frac{2e\lambda T_0\eta
P'_{r0}}{(1+2R_0\lambda T_0^2)^2}+\frac{r}{(1+2R_0\lambda
T_0^2)B_0^2}\left\{\left(\overset{(P_2)}{\psi_{rr}}-\overset{(P_2)}{\psi_{\phi\phi}}\right)\ddot{\eta}
+\dot{\eta}\overset{(P_1)}{\psi_{rr}}\right\}+\eta D_3=0,
\end{align}
where $X_1$ contains combinations of radial functions of $f(R,T)$
unusual relativistic cylindrical energy density and $D_3$ comprises
of corrections due to extra degrees of freedom of $f(R,T)$ gravity.
These quantities are given in Appendix \textbf{A}. The modified
versions of dynamical equations (\ref{52}) and (\ref{53}) would a
source of power stimulus in the examination of stability islands of
cylindrical self-gravitating interiors.

The boundary condition (\ref{27}) after using
Eq.(\ref{31})-(\ref{41}) give
\begin{align}\label{54}
P_{r0}^{\textrm{eff}}&\overset{\Sigma}=-\frac{\lambda
R_0^2T_0^2}{2}(1+2\lambda R_0T_0),\\\label{56}
\bar{P}_{r}&\overset{\Sigma}=\chi_4\eta.
\end{align}
where
\begin{align}\nonumber
\chi_4&\overset{\Sigma}=-e\lambda R_0T_0^2\{1+2\lambda
R_0T_0+\lambda R_0^3\}.
\end{align}
By making use of Eqs.(\ref{48}), (\ref{54}) and (\ref{56}), we get
\begin{equation}\label{57}
\delta_1\ddot{\eta}+\delta_2\dot{\eta}+\delta_3\eta\overset{\Sigma^{(e)}}=0,
\end{equation}
where
\begin{align}\nonumber
\delta_1=\left\{\overset{(P_2)}{\psi_{rr}}+\frac{(1+2 R_0\lambda
T_0^2)c}{rA_0^2}\right\}\frac{1}{(1+2R_0^2\lambda T_0)},~
\delta_2=\frac{\overset{(P_1)}{\psi_{11}}}{(1+2 R_0^2\lambda T_0)},~
\delta_3=\chi_4+\chi_2-\chi_3.
\end{align}
Equation (\ref{57}) has two independent different solutions in terms
of their graphical representations. One of these solutions
represents unstable while the other one describes the stable nature
of the collapsing cylindrical stellar object. In this manuscript, we
are concerned with the exploration of unstable phases of the compact
system in $f(R,T)$ gravity. So, we will restrict our analysis by
taking an oscillating solution. As seen from the perturbation scheme
that initially, our stellar model is in the phase of complete
hydrostatic equilibrium. For the examination of collapsing model, it
is required to limit radial perturbation parameters, i.e., $a,~b,~c$
and $e$, are all positive. Due to this, one can easily check that
$\omega_{\Sigma}^2>0$. In this chain, the required solution of the
above equation is
\begin{equation}\label{58}
\eta=-\exp{(\omega_{\Sigma}t)}, \quad \textmd{where} \quad
\omega_{\Sigma}=\frac{-\delta_2+\sqrt{\delta_2^2-4\delta_1\delta_3}}{2\delta_1}.
\end{equation}

\subsection{Hydrodynamical Equation}

In this section, we shall formulate $f(R,T)$ collapse equation which
would be helpful to check the instability regimes for our system.
For this purpose, we shall make use of  a familiar equation of state
(EoS), i.e., Harrison-Wheeler type EoS. It is useful to note that
this EoS peculiarly relate $\bar{\mu}$ with $\bar{P}_i$. This is
given as follows \cite{p32}
\begin{equation}\label{59}
\bar{P_i}=\Gamma_1\frac{P_{i0}}{P_{i0}+\mu_0}\bar{\mu}.
\end{equation}
where $\Gamma_1$ is known as adiabatic index. It measures the
fractional variations between pressure and energy density
experienced by matter configurations following the motion. In this
way, it reports how stiff a relativistic fluid is, thus suggesting
its alternative name, stiffness parameter. We assume $\Gamma_1$ to
be a constant entity throughout in our stability analysis of
relativistic stellar objects. In the scenario of N regime, the
instability of spherical self-gravitating systems depend purely on
the mean value of stiffness parameter \cite{p33}. However, in GR,
the stability relies not only on the average value of $\Gamma_1$ but
also on the star radius. In the study of relativistic
self-gravitating systems in modified gravity, the situation is quite
different. Chandrasekhar \cite{zf38} calculated a simple numeric
value, $\frac{4}{3}$, through $\Gamma_1$ for the stability of
relativistic isotropic sphere. But in modified gravity, finding of
such type of numeric value is impossible. The dark source terms
coming from modified gravity would greatly make ones calculations
cumbersome. Due to this reason, many researchers \cite{cc1} have
found such instability ranges through $\Gamma_1$ depending upon
radial configurations of matter variables. We are also expecting
such type of solution in our approach. Continuing in this way, the
integration of Eq.(\ref{52}) yields
\begin{equation}\label{60}
\bar{\mu}=\chi_1\eta,~\textmd{where}~\chi_1=-X_1.
\end{equation}
Using this value in Eq.(\ref{59}), we get
\begin{align}\label{61}
&\bar{\mu}-\bar{P}_r=\chi_1\left\{1-\frac{P_{r0}\Gamma_1}{(P_{r0}+\mu_0)}\right\}\eta,\\\label{62}
&\bar{P}_r-\bar{P}_\phi=\Gamma_1\chi_1\left\{\frac{P_{r0}}{P_{r0}+\mu_0}-\frac{P_{\phi0}}{P_{\phi0}+\mu_0}\right\}\eta
\end{align}
Using Eqs.(\ref{60})-(\ref{62}) in Eq.(\ref{53}), we obtain
\begin{align}\nonumber
&\Gamma_1\left(\frac{P_{r0}\chi_1}{P_{r0}+\mu_0}\right)'\frac{\eta}{(1+2R_0^2\lambda
T_0)}+\frac{\Gamma_1P_{r0}\chi_1\xi}{(P_{r0}+\mu_0)(1+2R_0^2\lambda
T_0)}\eta-\frac{A_0A_0'\chi_1\eta}{B_0^2}\\\label{63}
&+\left\{\frac{P_{r0}}{P_{r0}+\mu_0}-\frac{P_{\phi0}}{P_{\phi0}+\mu_0}\right\}
\Gamma_1\frac{r\chi_1}{B_0^2} \frac{(1+2R_0^2\lambda
T_0)}{(1+2R_0\lambda T_0^2)}\eta+\Phi\eta=0,
\end{align}
where
\begin{align}\nonumber
\Phi&=\frac{r}{(1+2R_0\lambda
T_0^2)B_0^2}\left\{\left(\overset{(P_2)}{\psi_{rr}}-\overset{(P_2)}{\psi_{\phi\phi}}\right)\omega^2
+\omega\overset{(P_1)}{\psi_{rr}}\right\}-\frac{A_0A'_0}{B_0^2(1+2R_0\lambda
T_0^2)}\left(\omega\overset{(P_1)}{\psi_{rr}}+\omega^2\overset{(P_2)}{\psi_{rr}}\right)\\\nonumber
&+\frac{2e\lambda T_0 P'_{r0}}{(1+2R_0\lambda
T_0^2)^2}+\frac{\chi_1'\eta}{(1+2R_0^2\lambda
T_0)}+\frac{2\chi_1\lambda \eta}{(1+2R_0^2\lambda
T_0)}\left[R_0(2T_0R'_0+R_0T'_0)-\frac{2R_0^2\lambda
T_0^2}{(1+2R_0^2\lambda T_0)}\right]\\\label{64} &+
D_3+\frac{2R_0^2\lambda T_0\chi_1'}{(1+2R_0^2\lambda
T_0)}\eta-2\lambda
R_0\chi_1\frac{(2T_0R'_0+T_0'R_0)}{(1+2R_0^2\lambda
T_0)}\eta,\\\nonumber \xi&=2\lambda
\left[(R_0T'_0+2T_0R'_0)R_0-(2R_0T'_0+T_0R'_0)T_0\frac{(1+2R_0^2\lambda
T_0)}{(1+2R_0\lambda
T_0^2)}\right]+2R_0\lambda(2R_0'T_0+T_0'R_0)\\\label{65}
&-\frac{A_0A'_0}{B_0^2}(1+2R_0^2\lambda T_0).
\end{align}
This is the hydrodynamical equation for our considered systems which
is framed in $f(R,T)$ gravity. It includes all the ingredients that
are required to study the unstable phases of the cylindrical
geometry which is coupled with locally anisotropic fluid
relativistic stellar matter.

\subsection{Newtonian Limit}

In order to assess N limit constraint, We consider
\begin{equation*}
\mu_0\gg P_{j0}, \quad A_0=1, \quad B_0=1.
\end{equation*}
Under this chain, the collapse equation (\ref{63}) provides
\begin{align}\label{66}
&\Gamma_1[(P_{r0}\chi_{1N})'+P_{r0}\chi_{1N}\xi_{N}+(P_{r0}-P_{\phi0})\chi_{1N}\zeta]\eta=|\Phi_N(1+2\lambda
R_0^2T_0)\eta|,
\end{align}
where $\zeta=\frac{1+2\lambda R_0^2T_0}{(1+2R_0T_0^2}$ and subscript
$N$ symbolizes that corresponding terms are approximated within N
epoch. Now, taking the value of $\eta$ from Eq.(\ref{58}), and
manipulating above equation, one can achieve above results
independent of temporal coordinate. The instability constraint for
cylindrically symmetric celestial object in $f(R,T)$ gravity is
obtained as follows
\begin{align}\label{67}
\Gamma_1<\frac{|\Phi_N(1+2\lambda R_0^2T_0)|}
{|(P_{r0}\chi_{1N})'|+|P_{r0}\chi_{1N}\xi_{N}|+|(P_{r0}-P_{\phi0})\chi_{1N}\zeta|}.
\end{align}
For cylindrical system coupled with isotropic pressure, we find
\begin{align}\label{68}
\Gamma_1<\frac{|\Phi_N(1+2\lambda R_0^2T_0)|}
{|(P_{0}\chi_{1N})'|+|P_{0}\chi_{1N}\xi_{N}|}.
\end{align}
The system will undergo in the stable phase against collapse if
modified gravitational forces induced by $|\Phi_N(1+2\lambda
R_0^2T_0)|$ are higher than that produced by
$|(P_{0}\chi_{1N})'|+|P_{0}\chi_{1N}\xi_{N}|$ , i.e., principal
stresses and non-attractive gravitational forces thereby providing
stability constraint $\Gamma_1>1$. For hydrostatic equilibrium, the
system would need to balance forces produced by $|\Phi_N(1+2\lambda
R_0^2T_0)|$ and $|(P_{0}\chi_{1N})'|+|P_{0}\chi_{1N}\xi_{N}|$. This
leads the system to reach at equilibrium condition i.e.,
$\Gamma_1=1$. On the other hand, if a stellar model has attained
forces coming from $|\Phi_N(1+2\lambda R_0^2T_0)|$ lesser than that
of $|(P_{0}\chi_{1N})'|+|P_{0}\chi_{1N}\xi_{N}|$, then the
relativistic interior will come in unstable window thus giving
$\Gamma_1\in(0,1)$. The quantities $\xi$ and $\chi_1$ contain
$f(R,T)$ corrections controlled by a parameter $\lambda$. Therefore,
higher order curvature terms have greatly modify the range of
dynamical instability of relativistic cylindrical object. It can be
seen from expressions (\ref{67}) and (\ref{68}) that anisotropy in
pressure along with dark source terms have greatly restrict the
instability regions thus producing difficulty for the system to
leave stable configurations against collapse.

\subsection{Post Newtonian Limit}

For instability conditions with pN approximation, we take
\begin{equation}\label{69} \mu_0\gg P_{j0},\quad
A_0=1-\frac{m_0\mathcal{G}}{r\mathcal{C}^2}, \quad
B_0=1+\frac{m_0\mathcal{G}}{r\mathcal{C}^2},
\end{equation}
where $\mathcal{G}$ and $\mathcal{C}$ are gravitational constant and
speed of light, respectively. It is worthy to note that we are
interested to consider effects upto $~O\left(\frac{m_0}{r}\right)$
in the following calculations. Equation (\ref{46}) provides
\begin{equation}\label{70}
\frac{A''_0}{A_0}=-\frac{A'_0B'_0}{A_0B_0}+\frac{B_0^2}{(1+2\lambda
R_0T_0^2)}\left[(\mu_0+P_{\phi0})(1+2\lambda R_0^2T_0)
+\frac{\lambda}{2}R_0^2T_0^2+\overset{(S)}{\psi_{22}}\right].
\end{equation}
By making use of Eqs. (\ref{44}) and (\ref{51}), we have
\begin{align}\label{71}
\frac{B'_0}{B_0}&=\frac{4m'_0}{(l-8m_0)},\\\label{72}
\frac{A'_0}{A_0}&=\frac{2r^2l(\mu_0+P_{r0})(1+2\lambda
R_0^2T_0)+\lambda R_0^2T_0^2r^2l-4(l-8m_0)\lambda T_0^2}
{2r(l-8m_0)(1+2\lambda R_0T_0^2+2\lambda
rT_0)}=\varphi\textmd{(say)}.
\end{align}
Using Eqs. (\ref{69})-(\ref{72}) and (\ref{58}), Eq.(\ref{63})
yields
\begin{align}\label{73}
&\frac{\Gamma_1\eta\Pi}{(1+2\lambda
R_0^2T_0)}=|\Phi_{pN}\eta|+\left|\left(1-\frac{4m_0}{r}\right)\varphi\chi_{1pN}\eta\right|,
\end{align}
where $\Pi=\left[\left(\frac{P_{r0}\chi_{1pN}}{\mu_0+P_{r0}}\right)'
+\frac{P_{r0}\chi_{1pN}\xi_{pN}}{\mu_0+P_{r0}}+\left\{\frac{P_{r0}}{\mu_0+P_{r0}}
-\frac{P_{\phi0}}{\mu_0+P_{\phi0}}\right\}\frac{\chi_{1pN}\zeta}{B_0^2}\right]$.
For the physical applicability of the above conditions, it is worthy
to constraint that all terms in the above are positive. The
instability constraint for locally anisotropic cylindrical system in
$f(R,T)$ is calculated as follows
\begin{align}\label{74}
\Gamma_1<\frac{|\Phi_{pN}(1+2\lambda
R_0^2T_0)|+\left|\left(1-\frac{4m_0}{r}\right)\varphi\chi_{1pN}(1+2\lambda
R_0^2T_0)\right|} {\Pi}.
\end{align}
For locally isotropic cylindrical systems, we reach the same
constraint as above with the difference that $\Pi$ will has the
following value
\begin{align}\label{75}
\Pi=\left(\frac{P_{0}\chi_{1pN}}{\mu_0+P_{0}}\right)'
+\frac{P_{0}\chi_{1pN}\xi_{pN}}{\mu_0+P_{0}}
\end{align}
The locally isotropic self-gravitating system will enter in the
unstable window, lest it satisfies the relation (\ref{74}) with
$\Pi$ given in (\ref{75}). This shows that pN unstable epochs relies
on combinations of principal stresses gradients, repulsive $f(R,T)$
gravity dark source terms and adiabatic index, $\Gamma_1$. These
terms then depends upon the static profiles of the corresponding
quantities. This asserts the importance of hydrostatic equilibrium
parameters. The quantities $\chi_{pN}$ and $\xi_{pN}$ are producing
gravitational effects due to dark source terms. The extra degrees of
freedom induced by $\lambda$ terms produce complexity in the
instability regions, thus delaying the collapse.

\section{Summary and Discussion}

The study of compact cylindrical object in modified gravity captures
many realistic features of the cosmos. The energy distribution of
universe also poses interesting puzzles to astrophysicists and
cosmologists. Compact celestial objects are an ideal natural
laboratory to look for observational signatures and possible
modifications in Einstein's gravity. There have been number of
researchers who have discussed the stable as well as unstable
regions of compact celestial objects within the framework of
Einstein's gravity. Some researchers have also examined these epochs
with higher order corrections in Einstein's theory. This paper
explores instability constraints for cylindrically symmetric
anisotropic celestial object in the background of fourth order
$f(R,T)$ gravity. The $f(R,T)$ theories yield corrections to the
field connected with the matter as compared to the usual GR field.
Many researchers used expansion-free scenario to keep the stiffness
parameter irrelevant in the study of dynamical instability. Here, we
focussed our analysis by encapsulating the effects predicted by
adiabatic index or stiffness parameter.

A very promising way to explain this is to keep the non-zero
contribution of expansion scalar during evolution of cylindrical
model. We explored couple of dynamical equations using contracted
form of Bianchi identities with $f(R,T)$ effective energy momentum
tensor. It is an established fact that the stars are in a state of hydrostatic
equilibrium due to the balance of two conflicting effects, namely,
the gravitational force and the opposite internal thermal pressure
in the star's interior provided by nuclear fusion of the elements.
When any of these forces overcome the other one then stability of
the star is affected and can make explosive events depending upon
the size of star. In other words, any galactic model may be stable
in one state and turns out to be unstable in later stage. The
critical aspect is to determine the dynamical instability via linear
perturbations, but one cannot find that up to what degree these
techniques can demonstrate the stability problem. The instability
issues have much significance as it is closely related with our
current understanding of black hole physics and future evolution
of the universe.

Due to the complicated framework of $f(R,T)$ theory, it would be
significant to gather the instability constraints under some
approximations. Therefore, we impose radial perturbation technique
on dynamical as well as $f(R,T)$ field equations. The expansion as
well as Ricci scalars are also perturbed using linear perturbations
on metric as well as material profiles. The static configurations of
the field and dynamical equations are constructed within the
environment of a viable $f(R,T)$ model. A specific $f(R,T)$ model is
used to investigate both qualitative and quantitative behavior of
inhomogeneous and unstable system when compared with GR field. For
non-zero expansion case, we use adiabatic index to construct
collapse equation for the study of instability epochs in N and pN
regimes. At N regime, we have a flat background metric while for pN
region, we have adopted some known profile of the metric functions.
The main findings from the collapse equation under both
approximations are summarized as follows:\\
(i) With a physical interpretation and understanding of the under
discussion model, we have found the significant dependence of the stiffness parameter
on the static profile of the system including the geometrical and
physical quantities, particularly, the extra curvature ingredients
coming due to $f(R,T)$ theory of gravity. We would like to mention
here that there exist some expansion-free dynamical instability
analysis in the literature that explored instability ranges at both
N and pN regimes independent of this adiabatic index (stiffness parameter) \cite{zf43,e1}.\\
(ii) It is observed that the anisotropy due to pressure gradients
makes increment in the unstable regions of cylindrical object. It
indicates that the cylindrical geometry will be stable against
fluctuations until it violates the inequalities obtained in
(\ref{67}) and (\ref{74}) for both N and pN limits.\\
(iii) The results investigated here are well consistent with those
already exist in the literature for GR \cite{zf33} and modified
gravity \cite{cc1} under some specific constraints.

It is significant to mention that adiabatic index have a particular
numerical value, i.e., $\frac{4}{3}$ and $1$ for isotropic spherical
and cylindrical celestial objects, respectively, in the framework of
GR. The instability ranges depend upon the choice of $f(R,T)$ gravity
models. One can have different instability regions for different
models, i.e., by changing the model, the results may differ from
the present one as different models correspond to different eras.
Finally, we mention that our all consequences correspond to that of GR under
$f(R,T)=R$ limit.

The modified gravity theories (like $f(R,T)$ gravity) provide
a cosmologically viable explanation for the current expansion in our cosmos
and have been the center of attention during the last few decades. The dynamical
equations in this modified gravity have contribution from mater parts,
consequently, one can have particular system of equations with every selection of $L_M$

In order to provide physically realistic $f(R,T)$ gravity theory, one has to study
well-consistent as well as physically acceptable models of gravity. Such models not only
obeys the constraints in relativistic background set by the solar system and terrestrial experiments
but also shed light over the current acceleration in our cosmos. Also, they satisfy the constraints
required for their theoretical viability. Any cosmological model in modified gravity should avoid instabilities
(Ostrogradski's instability, Dolgov-Kawasaki instability and
tachyons) and describe the exact cosmological dynamics. The stability analysis is a cornerstone in astrophysical
discussions on gravitational collapse. Haghani \emph{et al.} \cite{yan1} and Odintsov
and S\'{a}ez-G\'{o}mez \cite{yan2} commented that Dolgov-Kawasaki
instability in $f(R,T)$ gravity requires similar sort of limitations on the arbitrary function in the Lagrangian
as in $f(R)$ gravity

The following conditions should hold for physically realistic
$f(R)$ models \cite{1c1}:
\begin{itemize}
\item The positivity of $f_{R}(R)$ is needed to avoid the emergence of ghost state with $R>\tilde{R}$, here $\tilde{R}$
is the today value of the Ricci invariant. Ghost appears very often, while
describing the modified gravity theories that indicates DE as a source
behind current accelerating cosmos. This may emerge due to the
mysterious repulsive force between supermassive or
massive celestial objects at long distances. The condition of
keeping positive the effective gravitational constant, $G_{eff}=\frac{G}{1+2\lambda T^2 R}$, is also of great importance
for the attractive feature of
gravity.
\item The positivity of $f_{RR}(R)$ is developed to avoid the appearance of tachyons
(which is any hypothetical particle traveling faster than speed of
light) with $R>\tilde{R}$. One can consider the imaginary rest-mass so that moving mass must be real
as the moving mass of these particles is imaginary.
\end{itemize}
If $f(R)$ models do not satisfy these conditions, then it would be
regarded as unviable. In addition to the above mentioned limitations on the arbitrary function in the Lagrangian
of $f(R)$ gravity, we require $1+2\lambda T R^2>0$
for $G_{eff}>0$. So, for the physically acceptable $f(R,T)$ models, one needs to
satisfy the following constraints
\begin{align}\nonumber
1+2\lambda T^2 R>0,~1+2\lambda T R^2>0,~\lambda T^2>0,~R\geq\tilde{R}.
\end{align}

On the other hand, cylindrical systems have been surprising relativists
since Levi-Civita constructed its vacuum solution. They serve as a
natural tool to explore physics that lies behind the two independent
parameters in Levi-Civita spacetime and
in particular, the one that describes the Newtonian energy per unit
length looks the most elusive. The fact that there are two parameters
while in its counterpart, Newtonian theory, has only one parameter
looks a sufficient justification for deserving more research. Besides,
there has been renewed interest in cylindrically symmetric sources in
relation with different, classical and quantum, aspects of gravitation.
Such sources may serve as test bed for numerical relativity, quantum
gravity and for probing cosmic censorship and hoop conjecture, among
other important issues, and represent a natural tool to seek the physics
that lies behind the two independent parameters in Levi-Civita metric.

\vspace{0.3cm}

\section*{Acknowledgments}

This work was partially supported by the research funds provided
by University of the Punjab, Lahore-Pakistan through a research
project No. D/4112/Est.I in the fiscal year 2017-2018.

\vspace{0.3cm}

\renewcommand{\theequation}{A\arabic{equation}}
\setcounter{equation}{0}
\section*{Appendix A}

The dark source terms $D_0$ and $D_1$ are given as follows
\begin{align}\nonumber
D_0&=\frac{\psi_{01,1}}{A^2}-\frac{\psi_{01}}{A^2}\left(\frac{A'}{A}+\frac{B'}{B}+\frac{2AA'}{B^2}
-\frac{CC'}{B^2}\right)-\frac{\psi_{00}}{A^2f_R}(B\dot{B}+C\dot{C})-\frac{\psi_{11}}{A^2f_R}\\\label{A1}
&\times B\dot{B}-\frac{\psi_{22}}{A^2f_R}C\dot{C},
\\\nonumber
D_1&=\frac{\psi_{01,0}}{B^2}-\frac{\psi_{01}}{B^2}\left(\frac{\dot{A}}{A}+\frac{\dot{B}}{B}
+\frac{2B\dot{B}}{A^2}
+\frac{C\dot{C}}{A^2}\right)-\frac{AA'}{B^2f_R}(\psi_{00}+\psi_{11})+\frac{CC'}{B^2f_R}\\\label{A2}
&\times(\psi_{11}-\psi_{22})-\left\{\frac{f-Rf_R}{2}-\psi_{11}\right\}_{,1}.
\end{align}
The components of extra curvature terms $\chi_is$ are
\begin{align}\nonumber
\chi_2&=\frac{1}{B_0^2(1+2\lambda
R_0T_0^2)}\left\{\left(\frac{c}{r}\right)'\frac{B'_0}{B_0}
+\frac{1}{r}\left(\frac{b}{B_0}\right)'-\frac{c''}{r}\right\}-\left(\overset{(P)}{\psi_{00}}-e\lambda
R_0T_0^2\right)\\\nonumber &-\left(\mu_0-\lambda
R_0^2T_0^2+\overset{(S)}{\psi_{00}}\right)\left(\frac{2b}{B^2_0}+\frac{2\lambda
T_0^2}{(1+2\lambda R_0T_0^2)^3}\right),\\\nonumber
\chi_3&=\frac{1}{(1+2{\lambda}R_0^2T_0)}\left[(1+2{\lambda}R_0T_0^2)\left\{
\frac{A'_0}{A_0}\left(\frac{c}{r}\right)'\frac{1}{r}\left(\frac{a}{A_0}\right)\right\}
-\left\{\frac{\lambda}{2}{R_0^2}T_0^2+\overset{(S)}{\psi_{11}}\right.\right.\\\nonumber
&\left.\left.+(\mu_0+P_{r0})(1+2{\lambda}R_0^2T_0)
\right\}\left\{\frac{2b}{B_0^2} +2e\lambda
T_0^2\right\}-\left[2e\lambda R_0^2(\mu_0+P_{r0})+e\lambda
R_0T_0^2\right]\right],\\\nonumber
\chi_5&=-\frac{(1+2{\lambda}R_0T_0^2)}{A_0B_0^3(1+2{\lambda}R_0^2T_0)}
\left\{b'A'_0+a'B'_0-\frac{2b}{B_0}(A'_0B'_0)+B_0a''-bA''_0\right\}
-\left(2R_0P_{\phi0}\right.\\\nonumber
&+\left.2\mu_0R_0^2+T_0^2+\frac{\overset{(P)}{\psi_{22}}}{e\lambda
R_0}\right)\frac{e\lambda
R_0}{(1+2{\lambda}R_0^2T_0)}-(1+2{\lambda}R_0^2T_0)\left[
\frac{\lambda}{2}R_0^2T_0^2+\overset{(S)}{\psi_{22}}\right.\\\nonumber
&+\left.(\mu_0+P_{\phi0})
(1+2{\lambda}R_0^2T_0)\right]\left[\frac{b}{B_0}+\frac{a}{A_0}+\frac{2e\lambda
T_0^2}{(1+2{\lambda}R_0T_0^2)}\right].
\end{align}
The mathematical expressions mentioned in Eqs.(\ref{52}) and
(\ref{53}) are
\begin{align}\nonumber
X_1(r)&=\left[2\lambda T_0\mu_0\frac{(eT_0+2R_0z)}{(1+2\lambda
R_0T_0^2)}+\frac{(1+2\lambda R_0^2T_0)}{(1+2\lambda
R_0T_0^2)}(\mu_0+P_{r0})\left(\frac{c}{A_0^2}+\frac{bB_0}{A_0^2}\right)\right.\\\nonumber
&-\left.4\lambda\mu_0R_0\frac{(eT_0R_0z)}{(1+2\lambda
R_0^2T_0)}-\frac{z}{(1+2\lambda R_0^2T_0)}-\frac{1}{A_0^2(1+2\lambda
R_0T_0^2)}\left\{bB_0\overset{(S)}{\psi_{11}}\right.\right.\\\nonumber
&+\left.\left.
(bB_0+rc)\overset{(S)}{\psi_{00}}+rc\overset{(S)}{\psi_{22}}\right\}-\overset{(P_1)}{\psi_{01}}\frac{1}{A_0^2}
-\overset{(S)}{\psi_{01}}\frac{1}{A_0^2}\left(\frac{A'_0}{A_0}
+\frac{B'_0}{B_0}-\frac{r}{B_0^2}\right.\right.\\\label{A3}
&\left.\left.+\frac{2A_0A'_0}{B_0^2}\right)\right]\left[\frac{(1+2\lambda
R_0T_0^2)(1+2\lambda R_0^2T_0)}{(1+4\lambda
R_0^2T_0)}\right],\\\nonumber D_3&=\frac{2\lambda
R_0P_{r0}}{(1+2\lambda
R_0^2T_0)}(2T_0e'+2eR'_0+R_0z')-\frac{4e\lambda^2T_0P_{r0}}{(1+2\lambda
R_0^2T_0)^2}\left[P_{r0}(2T_0R'_0\right.\\\nonumber
&\left.+R_0T'_0)-T_0(T_0R'_0+2R_0T'_0)\frac{(1+2\lambda
R_0^2T_0)}{(1+2\lambda
R_0T_0^2)}\right]-\frac{4e\lambda^2T_0P_{r0}}{(1+2\lambda
R_0T_0^2)^2}(T_0R'_0\\\nonumber &+2R_0T'_0)\left\{\frac{(1+2\lambda
R_0^2T_0)}{(1+2\lambda R_0T_0^2)}-R_0^2\right\}+\frac{2\lambda
T_0P_{r0}}{(1+2\lambda
R_0T_0^2)^2}\left(T_0e'+2eR_0\frac{T'_0}{T_0}\right.\\\nonumber
&\left.+2R_0z'\right)(1+2\lambda
R_0^2T_0)-(\mu_0+P_{r0})\frac{A_0A'_0}{B_0^2(1+2\lambda
R_0T_0^2)}\left\{\frac{a}{A_0}+\frac{a'}{A'_0}-\frac{2b}{B_0}\right.\\\nonumber
&\left.-\frac{2e\lambda T_0}{(1+2\lambda
R_0T_0^2)}\right\}(1+2\lambda R_0^2T_0)-\frac{2e\lambda
T_0\mu'_0}{(1+2\lambda R_0T_0^2)^2}-\frac{2e\lambda
R_0^2A_0A'_0}{B_0^2(1+2\lambda R_0^2T_0)}\\\nonumber
&\times(\mu_0+P_{r0})+\frac{4e\lambda^2 T_0P_{r0}}{(1+2\lambda
R_0T_0^2)^2}\left[2\lambda
R_0^2T_0^2\frac{(T_0R'_0+2R_0T'_0)}{(1+2\lambda
R_0T_0^2)}-\mu_0(2T_0R'_0\right.\\\nonumber
&\left.+R_0T'_0)\right]+2\lambda
R_0\mu_0\frac{(2T_0e'+2eR'_0+R_0z')}{(1+2\lambda
R_0T_0^2)}+\frac{4\lambda^2T_0^2R_0^2\mu_0}{(1+2\lambda
R_0T_0^2)^2}\left(2R_0z'+T_0e'\right.\\\nonumber
&\left.+2eR_0\frac{T'_0}{T_0}\right)-4e\lambda^2T_0\mu_0R_0^2
\frac{(T_0R'_0+2R_0T'_0)}{(1+2\lambda
R_0T_0^2)^2}\left\{\frac{2\lambda T_0}{(1+2\lambda
R_0T_0^2)}-R_0^2\right\}\\\nonumber & + \frac{2e\lambda
R_0^2}{(1+2\lambda R_0^2T_0)}\left(\mu'_0+\frac{T'_0}{2}\right)
\left[1-\frac{2\lambda R_0^2T_0}{(1+2\lambda R_0^2T_0)}\right]
+\frac{2\lambda R_0^2T_0\bar{\mu}'}{(1+2\lambda
R_0^2T_0)}\\\nonumber &-\frac{\lambda R_0^2 T_0z'}{(1+2\lambda
R_0^2T_0)}+\frac{2\lambda R_0}{(1+2\lambda
R_0^2T_0)}(\mu_0-P_{r0})\left[\frac{2e\lambda R_0^2}{(1+2\lambda
R_0^2T_0)}(2T_0R'_0\right.\\\nonumber
&+\left.R_0T'_0)-2T_0e'-2eR'_0-R_0z'\right]+r\frac{(P_{r0}-P_{\phi0})}{B_0^2(1+2\lambda
R_0T_0^2)}(1+2\lambda R_0^2T_0)\left\{\frac{c}{r}\right.\\\nonumber
&\left.+c'-\frac{2e\lambda T_0}{(1+2\lambda
R_0T_0^2)}-\frac{2b}{B_0}\right\}-\frac{2er\lambda
R_0^2}{(1+2\lambda
R_0T_0^2)}(P_{r0}-P_{\phi0})-\left\{\left(\overset{(S)}{\psi_{00}}\right.\right.\\\nonumber
&\left.\left.+\overset{(S)}{\psi_{11}}\right)\left(\frac{a}{A_0}+\frac{a'}{A'_0}-\frac{2b}{B_0}
-\frac{2e\lambda T_0}{(1+2\lambda
R_0T_0^2)}\right)+\overset{(P)}{\psi_{00}}+\overset{(P)}{\psi_{11}}\right\}\frac{A_0A'_0}{B_0^2(1+2\lambda
R_0T_0^2)}\\\label{A4} &\lambda(eR_0
T-0^2)'+\overset{(P)}{\psi_{00,1}}+\frac{r}{B_0^2(1+2\lambda
R_0T_0^2)}\left(\overset{(P)}{\psi_{11}}-\overset{(P)}{\psi_{22}}\right).
\end{align}

\end{document}